\documentclass{article}
\usepackage{graphicx}

\title{Feynman rules for Coulomb gauge QCD}
\author{A. Andra\v si $^a$, J. C. Taylor $^b$}

\begin{document}

\maketitle

{\it a Rudjer Bo\v skovi\' c Institute, Zagreb, Croatia}

{\it b Department of Applied Mathematics and Theoretical Physics, University of Cambridge, Cambridge, UK}


\def\p{\partial}
\def\X{{\bf X}}
\def\P{{\bf P}}
\def\R{{\bf R}}
\def\Q{{\bf Q}}
\def\K{{\bf K}}
\def\A{{\bf A}}
\def\Z{{\bf Z}}
\def\Y{{\bf Y}}
\def\e{\epsilon}
\section*{Abstract}
The Coulomb gauge in nonabelian gauge theories is attractive in principle, but beset with
technical difficulties in perturbation theory. In addition to ordinary Feynman integrals, there
are, at 2-loop order, Christ-Lee  (CL) terms, derived either by correctly ordering the operators in the Hamiltonian,
or by resolving ambiguous Feynman integrals. Renormalization theory depends on the sub-graph
structure of ordinary Feynman graphs. The CL terms do not have a sub-graph sructure. We show how to 
carry out renormalization in the presence  of CL terms, by re-expressing these as `pseudo-Feynman' integrals. We also explain how energy divergences cancel.

\section{Introduction}

There are some reasons to be interested in the Coulomb gauge in QCD. It is the only explicitly
unitary gauge (we discount axial gauges, which have undefined denominators $1/(n.k)$\footnote{We include here the temporal gauge, in which $n$ is time-like, although this gauge is sometimes taken as a starting point.}).
It may be useful for bound-state problems. It has been used in lattice simulations, and
it has been made the basis for a discussion of confinement \cite{zwanziger}, \cite{cuch}, \cite{scz}, 
\cite{szi}, \cite{alk}, \cite{nak}, \cite{rei}, \cite{lan}, \cite{naka}, \cite{epple}, \cite{burgio},\cite{camp} (the existence of a Hamiltonian allows the
use of a variational principle \cite{rce}).
But there are complications in formulating correct Feynman rules, all concerned with
the convergence of integrals over the energy-components of internal energy-momentum
variables. These problems are eased by using the Hamitonian (phase-space) formulation,
in which the electric field ${\bf E}^a$ is one of the dynamical variables \cite{mohapatra}.
The Lagrangian in the Hamiltonian formalism, and the Feynman rules are stated in Appendix A.

Even with this formulation, divergent energy integrals remain in individual Feynman graphs,
which are cancelled when suitable sets of graphs are combined. This has been illustrated
in \cite{AAcanc}.

A more subtle complication about the energy integrals remains. This was first recognized as
a question of operator ordering by \cite{christlee} (which contains references to earlier
related work, and see also \cite{cheng1}). But an equivalent approach is to regard it as coming from ambiguous
double energy integrals at 2-loop order (see \cite{cheng2}). This integral is (we use bold capital letters for the space components
of 4-vectors)
$$\int dp_0 dq_0 {p_0\over p_0^2-{\bf P}^2+i\epsilon}{q_0\over q_0^2-{\bf Q}^2+i\epsilon} \eqno(1.1)$$
This double integral is not absolutely convergent, and might  be defined as a repeated integral
in two different ways:
$$\int dp_0{p_0\over p_0^2-{\bf P}^2+i\epsilon}\int dq_0{q_0\over q_0^2-{\bf Q}^2+i\epsilon}=0 \eqno(1.2)$$
or
$$\int { dr_0\over 2\pi}\int {dq_0\over 2\pi}{-i(r_0+q_0)\over (r_0+q_0)^2-{\bf P}^2+i\epsilon}{iq_0\over q_0^2-{\bf Q}^2+i\epsilon}=1/4.\eqno(1.3)$$

The above ambiguous integrals occur in 2-loop order in graphs containing just two propagators $\langle 0|T(A_i^a(x)E_j^b(y))|0\rangle$ (our notation is explained in Appendix A) and a connected chain
of Coulomb propagators. 

 Christ and Lee \cite{christlee} (and previous authors referenced there) studied the correct ordering of operators (including the Fadeev-Popov determinant) in the Hamiltonian. They found two new terms called $V_1$ and $V_2$ (of order $\hbar^2$) which had to
 be added. At the same time, the ordinary parts of the Hamiltonian were to be Weyl ordered
 (that is a certain average over different orderings). It was argued that this Weyl ordering removed all the ambiguous integrals like (1.1), their place being taken by $V_1$ and $V_2$.
  \footnote{In \cite{christlee},
 (section VI) it is stated that the integral (1.7) is zero. There seems to be no explicit statement about
 integrals like (1.1).  If the Weyl ordering implies that all integrals like (1.1) are zero, there seems to lead to a contradiction because of the identity (1.4).}

An alternative approach was given in \cite{doust}. There, the ambiguity of (1.1) was
resolved by showing that suitable combinations of Feynman graphs lead to the unambiguous, absolutely convergent, 
combination (defining $p_0+q_0+r_0=0$)
$$\int {dp_0\over 2\pi}{ dq_0\over 2\pi} \Big({ip_0\over p_0^2-{\bf P}^2+i\epsilon}{iq_0\over q_0^2-{\bf Q}^2+i\epsilon} +
{iq_0\over q_0^2-{\bf Q}^2+i\epsilon}{ir_0\over r_0^2-{\bf R}^2+i\epsilon} $$ $$+
{ir_0\over r_0^2-{\bf R}^2+i\epsilon}{ip_0\over p_0^2-{\bf P}^2+i\epsilon}\Big)=1/4 .\eqno(1.4)$$
This double integral may be done in any order, and therefore confirmed by using (1.2) and (1.3).

Equation (1.4) is related to the identity
$$\e(t_1-t_2)\e(t_2-t_3)+\e(t_2-t_3)\e(t_3-t_1)+\e(t_3-t_1)\e(t_1-t_2)=-1,$$
which has an unambiguous limit as the times tend to zero.

Actually, (1.4) is not quite sufficient to resolve all ambguity. In \cite{doust} one further
rule was used. This is that integrals of the form 
$$X\equiv \int dq_0 d{\bf Q} dr_0 d{\bf R}{q_0\over q_0^2-{\bf Q}^2+i\epsilon}{r_0\over r_0^2-{\bf R}^2+i\epsilon} f({\bf Q})g({\bf R})=0 \eqno(1.5)$$
where $f,g$ are any two functions. This rule could not be obtained from within the
Coulomb gauge, but was deduced from the limit of a gauge which interpolates between
the Feynman gauge and the Coulomb gauge. The argument is that, in the interpolating gauge, the
double integrand still factorizes, taking the form
$$\int dq_0 d^3\Q dr_0 d^3\R  F(\Q,q_0^2)G(\R,r_0^2),\eqno(1.6)$$
where each energy integral is now convergent, and so is unambiguously zero.\footnote{There is no
actual Feynman graph giving an integral like (1.5). There are (non-zero) Feynman graphs which give integrals
like (1.5) but with $r_0$ replaced by $p_0$ (as in Fig.8(c)), and (1.5) is just something which is added on to these
in order to get the combination in (1.4).}

From the identity (1.4) and the rule (1.5), it was shown (provided that the external gluons are transverse) in \cite{doust} that the additions
$V_1+V_2$ of \cite{christlee} are contained within ordinary Feynman integrals.

The problem with the above treatments of the ambiguous integrals is as follows. It is an important
property of Feynman graphs that, as a simple example, a 1-loop graph may appear as a sub-graph
of a 2-loop graph; and when it does it represents the same integral in both cases.
This property is vital to the process of renormalization of ultra-violet (UV) divergences.
The necessary counter-terms are found from 1-loop integrals, and are then used to
cancel the sub-divergences at 2-loops.

In the presence of the ambiguous integrals (1.1), this property is not obvious. For example,
the 1-loop graph in Fig.5(c) contains the integral
$$\int dp_0{p_0\over p_0^2-{\bf P}^2+i\epsilon} \eqno(1.7)$$
which is naturally, and correctly, taken to be zero. But in the 2-loop graph Fig.6(a),
which contains Fig.5(c) as a sub-graph, the ambiguous integral (1.1) appears, which according to
\cite{doust} has to be combined with other graphs to give (1.4). It is not obviously correct
to attach the value zero to Fig.5(c) in isolation. What is more, if the sub-graph in Fig.6(a)
is not zero, it is UV divergent, but the calculation of Fig.5(c) provides no counter-term
to cancel any sub-divergence in Fig.6(a). In a similar way, the 1-loop graph in Fig.5(a)
is unambiguous (and UV divergent), but it appears as a sub-graph in Fig.6(b) which contains the
ambiguous double integral (1.1).

This is a rather trivial example, but in the following we will encounter several similar cases.

It is the purpose of this paper to propose a solution to the dilemma.

We first have to remind the reader what the functionals $V_1$ and $V_2$ in \cite{christlee}
are.

\section{The functionals of Christ and Lee}

In order to state what these are, we first need to define the ghost propagator, $G$.
It is defined to be the solution of the equation
$$(\partial/ \partial X_i)\big(D_i^{ab}({\bf X})G^{bc}({\bf X},{\bf X}';{\bf A})\big)=\delta^{ac}\delta^3({\bf X}-{\bf X}'). \eqno(2.1)$$
Here $i,j,..$ are 3-vector indices, $a,b,..$ are colour indices,   we write a 4-vector $x_{\mu}=(t,X_i)$,
and $D^{ab}_i$ is defined in (A3).
In (2.1), $A_i^a(x)$ is an external gluon field which is transverse, $\partial A_i^a(x)/\partial X_i=0$.
$G$ is a functional of $A_i^a$.
Note that (2.1) is an instantaneous equation, there are no time derivatives. It is convenient to
define also
$$G_i^{ab}(\X,\X';\A)\equiv (\p/\p X_i)G^{ab}(\X,\X';\A). \eqno(2.2)$$

The zeroth order term in the perturbation expansion of $G$ is just the Coulomb potential
$$G^{ab}=\delta^{ab}{1\over 4\pi}{1\over |{\bf X}-{\bf X'}|}+O(g)\equiv\delta^{ab} G_0({\bf X}-{\bf X}')+O(g). \eqno(2.3)$$

In terms of the ghost propagator, the complete Coulomb potential operator is
$$C^{ab}(\mathbf{X},\mathbf{X}';\mathbf{A})=\int d^3\mathbf{X}''G_i^{ca}(\mathbf{X}'',\mathbf{X};\mathbf{A})G_i^{cb}(\mathbf{X}'',\mathbf{X'};\mathbf{A}). \eqno(2.4)$$

We need also to define the transverse projection operator
$$T_{ij}(\mathbf{X},\mathbf{Y})=\delta_{ij}\delta^3(\mathbf{X}-\mathbf{Y})-\p_i \p_jG_0(\mathbf{X}-\mathbf{Y}) .\eqno(2.5)$$

With this notation, the first Christ-Lee conribution to the Hamiltonian  is
$$V_1=-(g^2/8)f^{abc}f^{a'b'c'}\int d^3\mathbf{X}d^3\mathbf{Y}$$
$$\times G_i^{aa'}(\mathbf{X},\mathbf{Y};\mathbf{A})\delta^3(\mathbf{X}-\mathbf{Y})\delta^{bb'}
G_i^{c'c}(\mathbf{Y,}\mathbf{X};\mathbf{A})\eqno(2.6)$$

For the second, we first define
$$L_{ij}^{ab}(\mathbf{X},\mathbf{Y};\mathbf{A})\equiv \delta^{ab}T^{ij}(\mathbf{X},\mathbf{Y})+g\int d^3 \mathbf{Z}G_i^{ac}(\mathbf{X},\mathbf{Z};\mathbf{A})f^{cbe}A^e_k(\Z)T_{kj}(\mathbf{Z},\mathbf{Y})
\eqno(2.7)$$

Then
$$V_2=-(g^2/8)f^{abc}f^{a'b'c'}\int d\mathbf{X}d\mathbf{Y} L_{ij}^{aa'}(\mathbf{X},\mathbf{Y};\mathbf{A})
C^{bb'}(\mathbf{X},\mathbf{Y};\mathbf{A)}L_{ji}^{c'c}(\mathbf{Y},\mathbf{X};\mathbf{A}).\eqno(2.8)$$
Here we have not used the original form in \cite{christlee} but an equivalent given by \cite{doust}
in his equation (4.5.3).

The contributions to (2.6) and (2.8) in a perturbation series may be represented by 3-dimensonal Feynman graphs, in terms of the Fourier transform $A^a_i(\mathbf{K})$ of $A_i^a(\mathbf{X})$
(which we assume to satisfy ${\bf K}.{\bf A}^a=0$ \footnote{This condition is assumed also in the work of Christ and Lee. Thus, unlike the contributions form ordinary Feynman diagrams, the CL part of the effective action is not known for non-transverse external gluon fields.}). We shall call these CL graphs.
Typical examples, to order $g^6$ are shown in Fig.1(a) for $V_1$ and Fig.1(b) for $V_2$. Our graphical notation is explained in Appendix A. The rules are like those for a Euclidean field theory in 3-dimensions.

\begin{figure}
\centering
\includegraphics[width=11cm]{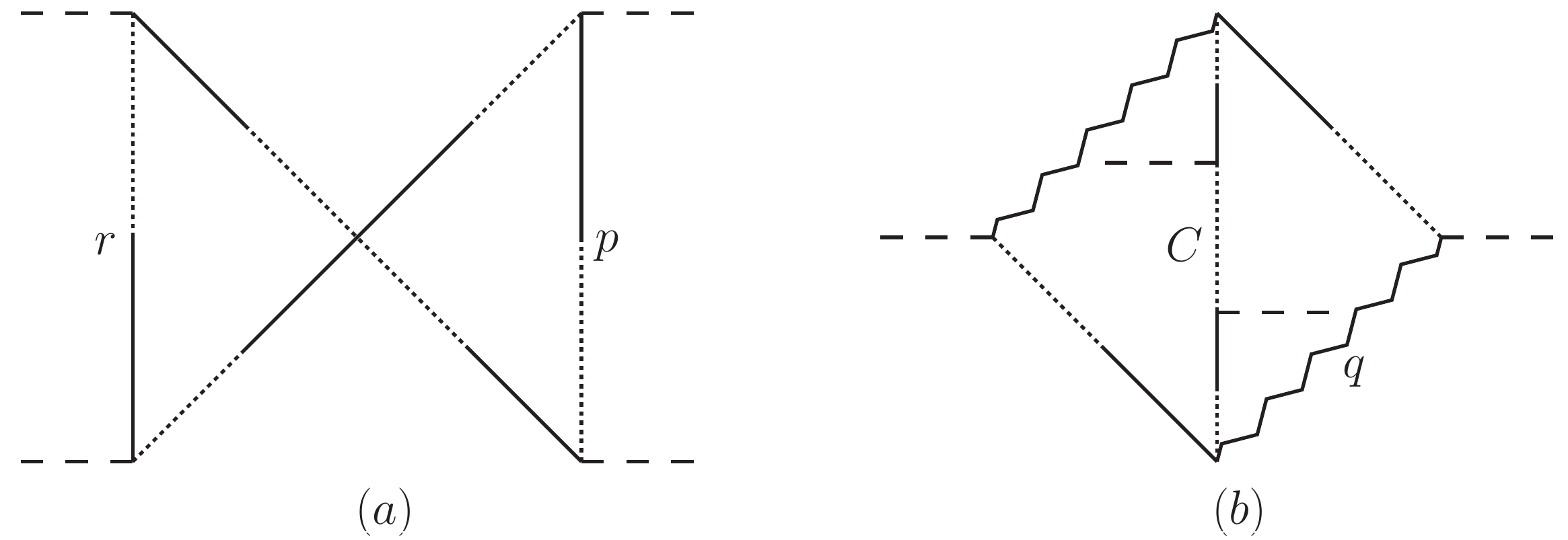}
\caption{Three-dimensional  graphs of order $g^6$ in momentum space. (a)  an example of a contribution
to $V_1$ in (2.6). (b) an example of a contribution to $V_2$ in (2.8). The central chain of lines is from 
$C$ in (2.4) and we call this the $C$-chain. The two
side chains are each from $L$. The zig-zag line denotes  (the Fourier transform of) $T_{ij}$.  The symbol  $C$ on the central segment in (b) draws attention to the Coulomb propagator
$G_0$ in (2.3), which in this case could be in any one of three positions on the central vertical
 $C$-chain, giving three equal contributions.}
\end{figure}

It will be useful to have a notation for the CL terms in momentum space, like the examples
in Fig.1. We will denote these integrals by $V_1(N)$ and $V_2(N)$ where $N$ is the total number
of external gluon lines. Then
$$-V_1(N)={g^{N+2}\over (2\pi)^6}f^{abc}f^{a'bc'}\int d^3\P d^3\R\sum_{n,n'}\tilde{G}_i^{aa'}(\P,\{n\})\tilde{G}_i^{c'c}(\R,\{n'\})$$
$$\equiv {1\over (2\pi)^6}\int d^3\P d^3\R W^N_1(\P,\R), \eqno(2.9)$$
where ${n}$ stands for the quantum numbers of $n$ external gluons, that is for $\{\K_\nu,i_\nu,a_\nu\}$
with $\nu=1,..,n$ and $n+n'=N$. Thus Fig.1(a) is an example with $n=n'=2$. We have put the minus sign
on the left in (2.9) because $-V_{1,2}$ are the contributions to the effective action.)

\pagebreak

 Similarly,
$$-V_2(N)={g^{N+2}\over (2\pi)^6}f^{abc}f^{a'b'c'}$$
$$\times \int d^3\P  d^3\Q d^3\R\delta^3(\P +\Q +\R)\sum_{n,n',n''}(n''+1)\tilde{L}_{ij}^{aa'}(\P,\{n\})\tilde{C}^{bb'}(\R,\{n''\})\tilde{L}_{ji}^{cc'}(\Q,\{n'\})$$
$$\equiv {1\over (2\pi)^6} \int d^3\P d^3\Q  W_2(\P,\Q,\R), \eqno(2.10)$$
where here $n+n'+n'' =N$. Fig.1(b) is an example with $n=n'=1, n''=2$. The factor $(n''+1)$ is inserted into 
(2.10) to take account of the $n''+1$ possible positions of the zero-order Coulomb propagator (2.3)
within $\tilde{C}$ (as indicated by the `$C$' in Fig.1(b).) In (2.10), $\P$ is defined to be the momentum through the transverse projector $\{\delta_{ij}-(P_iP_j/\P^2)\}$ in $L(\P)$, and $\Q$ likewise.
$\R$ is defined so that $\P +\Q +\R =0$.

The theorem proved in \cite{doust} may be stated as follows:
the sum of the ambiguous Feynman integrals (containing (1.1)) may be brought to the form (after some manipulation)

$$\Bigg[\int {d^4p\over (2\pi)^4}{d^4q\over (2\pi)^4 }\Big({p_0\over p_0^2-{\bf P}^2+i\epsilon}{q_0\over q_0^2-{\bf Q}^2+i\epsilon} +
{r_0\over r_0^2-\R^2+i\epsilon}{p_0\over p_0^2- \P^2+i\epsilon}$$
$$+{r_0\over r_0^2-{\bf R}^2+i\epsilon}{q_0\over q_0^2-{\bf Q}^2+i\epsilon}\Big)(W^N_1+W^N_2)\Bigg]
+X\eqno(2.11)$$
where $W^N_1$ and $W^N_2$ are defined in (2.9) and (2.10) and $X$ is an integral of the form of  (1.5), and which is taken to be zero.
The integrals in (2.11) are 4-dimensional; but not all the terms are Feynman integrals
generated from the rules in the appendix, because the denominator in (1.7)
can appear without the transverse projector $\{\delta_{ij}-(P_iP_j/\P ^2)\}$ in the numerator.
Also, since (1.4) is independent of $\P, \Q, \R$, the identification of these with the integation
variables in (2.9) and (2.10) is arbitrary.

If dimensional regularization is used, with $3-\epsilon$ spatial dimensions, the 1-loop sub-integrations in  CL graphs like those in Fig.1 have no divergences, that is no poles at  $\epsilon=0$. The complete
2-loop integrals are UV divergent, with single poles at $\epsilon=0$. At first sight, it is unexpected
that there are no divergent sub-graphs in CL graphs. But these sub-graphs have no meaning
on there own; there are no 1-loop CL graphs.

As explained in section 1, the problem is how to combine the CL graphs
with  ordinary Feynman graphs, particularly for the purpose of renormalization.
In the next section, we propose an answer to this dilemma.
\footnote{This problem is briefly alluded to in \cite{christlee} at te end of section VI. It is noted that
calculation of
an order $g^4$ contribution to $V_1$ involves singular terms of the form $\delta^3(\X)/|\X|$. It is suggested that
these may be 'relevant to the cancellation of divergences from the usual two-loop  Feynman graphs'.
Using dimensional regularization, we do not encounter such singular terms.}

\section{Expressing Christ-Lee terms as Feynman-like integrals}

We want to make (2.11)  more like ordinary Feynman integrals. For the $W^N_2$ term this
is easy, but for the $W^N_1$ term it can only be achieved partially.

The order of the $p_0, q_0$ integrations in (2.11) is optional, because of the identity (1.4).
Let us choose the order to be that in (1.3). Then only the first term in the square bracket (2.11)
contributes; the other two terms are zero because of equations like (1.2). Then,
in the contribution from $W^N_2$, there appears the correct Feynman propagtor
$$\left(\delta_{ij}-{P_iP_j \over \P^2}\right){p_0\over p_0^2-\P^2+i\epsilon},\eqno(3.1)$$
and similarly for $\Q$. An example is shown in Fig.2(a).  This looks like an ordinary Feynman graph,
but it is supplemented with the rule that the ambiguous integral (1.1) is in this case to be
interpreted in the order in (1.3). Thus we are to some extent reversing the process
carried out in \cite{doust}. But now we attribute the value of (1.1) to a single type of graph,
rather than to the combination in (2.11). The instantaneous propagators (denoted by dotted lines)
are the same in the  graphs in Fig.2 as in  the CL graph in Fig.1(b). 
We will call graphs like Fig.2(a) 'pseudo-Feynman graphs'.

The above process does not generate any graph like the one in Fig.2(b), although this looks like an
ambiguous Feynman graph. Thus we may say that pseudo-Feynman graphs like Fig.2(b) are zero,

Note also that the above rule is consistent with the term $X$ in (2.11) being zero (because of (1.2)).

 The general rule is that
the energy integrals are to be done in the order with the energy on the $C$-chain of lines (defined to be
a factor  like $C$ as defined in (2.4)) last.
The $C$-chain is the central one in Fig.2(a) but the right hand chain in Fig.2(b) (it is the chain
which contains a segment with just the Coulomb propagator $1/{\bf K}^2$)). However, there
are some special exceptions to this second rule, arising form $V_1$.

\begin{figure}
   \centering
   \includegraphics[width=11cm]{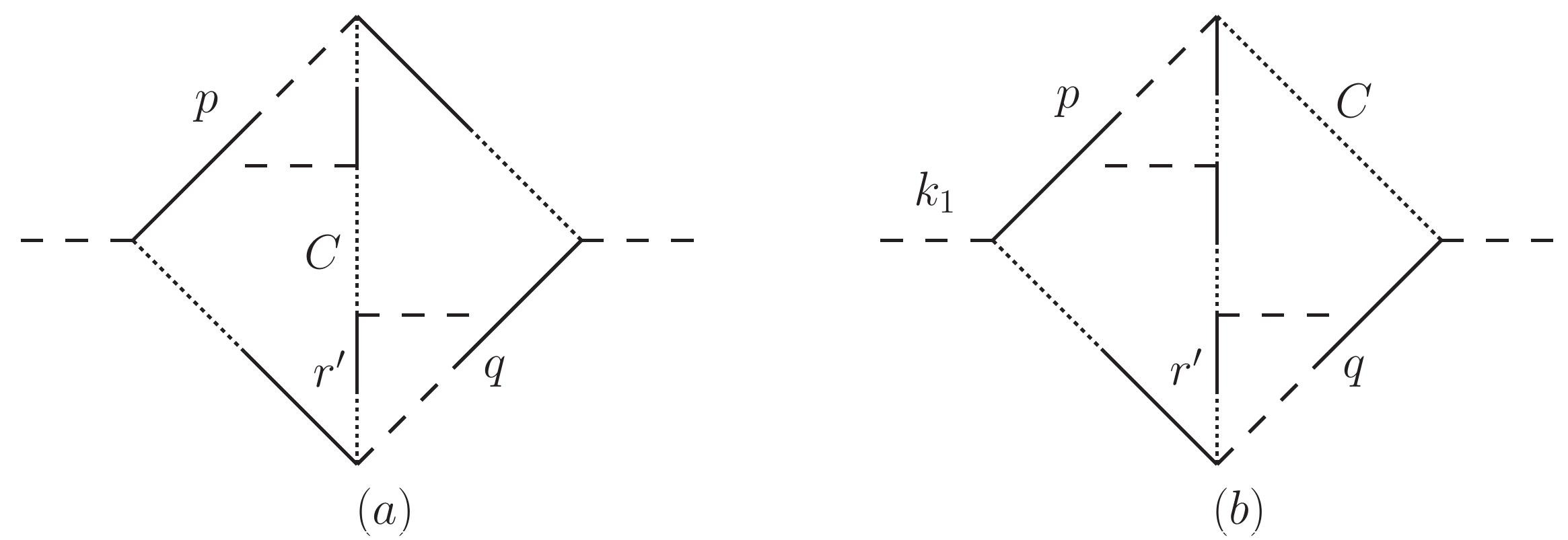} 
   \caption{(a) is  a pseudo-Feynman graph which is equivalent to the CL graph in Fig.1(b), provided that the
   energy integrals in the inserted factors are done in the order in (1.3). (b) is a Feynman graph of a similar
   type, but the $C$-chain (which in this case reduces to just a single line) is now the one on the top right, and the energy integrals are to be done in
   the order of (1.2), consequently giving a zero contribution. In this graph, $r'=r+k_1$ so that $p+q+r=0$.
   The symbol $C$ draws attention to the propagator $G_0$ in (2.3). In (a), this segment can be in any of three places, giving three equal contributions.}
   \end{figure}

We have not been able to express all of the  $W^N_1$ in  (2.11) in terms of pseudo-Feynman graphs, but just some special parts of $W^N_1$. But these will turn out to be sufficient to treat renormalization. Fig.3(a) shows an example
of a class of diagrams of the form of the CL graphs in Fig.1(a), which have any number of
external gluon lines on the left, but only one on the right. The CL integral has the form
$${g^{3+n''}\over 4}\int {d^3\R\over (2\pi)^3}  F_k(\R, \K_j) \int {d^3\Q\over (2\pi)^3} {Q_k\over \Q^2}{Q'_i\over \Q'^2}  \eqno(3.2)$$
where the details of the function $F$ are not important (and $n''$ is the number of external gluons in $F$). Because the external gluon on the right
is transverse, we can replace $Q'_i$ by $-Q_i$.  We can also insert a term $\delta_{ik}\int d^3\Q (1/\Q'^2)$
which is zero with dimensional regularization. Thus we get instead of (3.2)
$$(1/4)\int {d^3 \R\over (2\pi)^3} F_l(\R, \K_l) \delta_{lk}\int {d^3\Q\over (2\pi)^3} \left(\delta_{ik}-{Q_iQ_k\over \Q^2}\right){1\over \Q'^2} . \eqno(3.3)$$

Now we insert the integrand in  (3.3) into (2.11), and obtain
$$(2\pi)^{-8}\int d^4rd^4pF_l(\R,\K_l){\delta_{lk}p_0\over p_0^2-\P^2}\left(\delta_{ik}-{Q_iQ_k\over \Q^2}\right) {q_0\over q_0^2-\Q^2}{1\over Q'^2} \eqno(3.4)$$
where the $p_0$ and $r_0$ integrals are to be done in that order.
(3.4) is almost a Feynman integral, except that
the $p$-propagator has only the numerator $\delta_{lk}$ not the transverse projection operator in (3.1).
  This incomplete Feynman integral is denoted by the graph in Fig.3(b), where the
cross bar on the $p$-line denotes its incomplete numerator. Apart from this feature, Fig.3(b)
is a particular case of the class  in Fig,2(b). Thus there are special exceptions to the statement
that the terms in Fig.2(b) are zero.

(Colour factors are omitted in (3.3) and (3.4).)

\begin{figure}
\centering
\includegraphics[width=11cm]{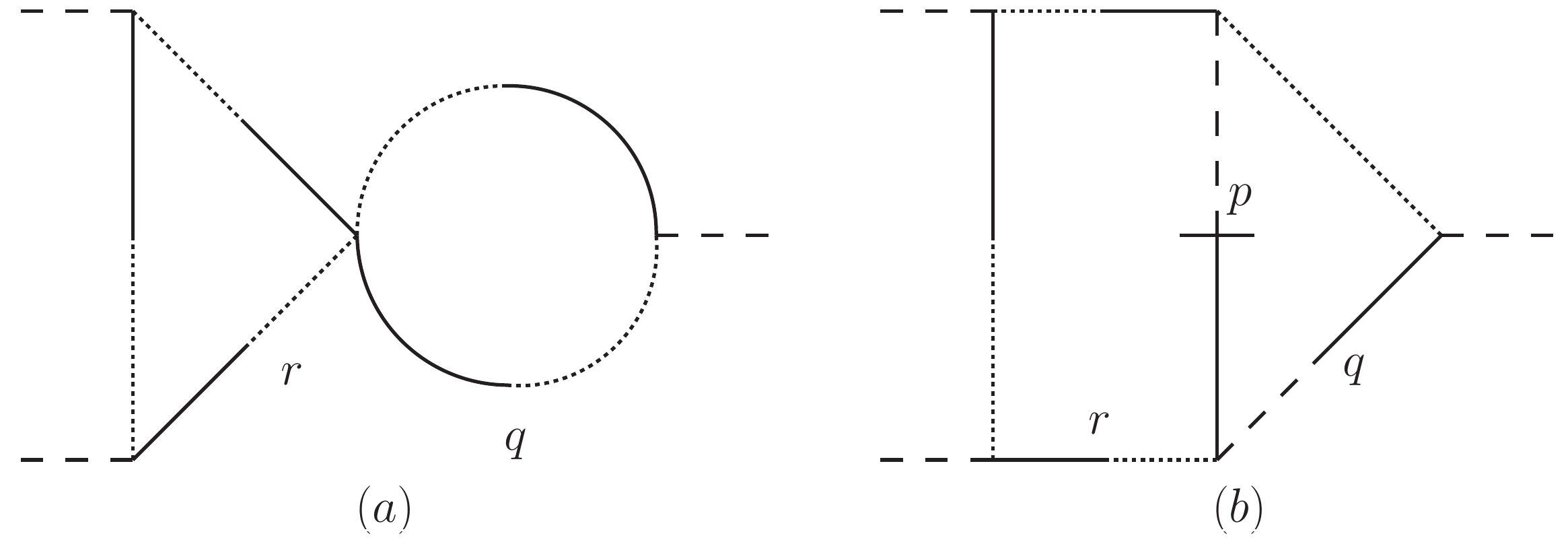}
\caption{(a) A particular class of CL diagrams from Fig.1(a), with only one external gluon on the
right. (b) An incomplete pesudo-Feynman diagram which gives something equivalent to (a). The cross
bar on the $p$-line denotes the incomplete numerator.}
\end{figure}

\section{Renormalization}

Renormalization counter-terms are derived from 1-loop integrals. These have to cancel
divergences in sub-graphs at 2-loops. The problem is how to identify these sub-graphs
when there are CL terms which do not have the sub-graph structure of normal Feynman graphs.
We will show that, re-expressing the CL integrals as pseudo-Feynman graphs, like those
in Figs.2(a) and 3(b) enable us to solve this problem.

The divergent part of 1-loop effective action is known, for example in \cite{AAdiv}: 
$$ -C_G{g^2\over 16\pi^2}{1\over \epsilon}\Big[-{11\over 12}{F^{a}_{ij}}^2-{1\over 6}{F^a_{0i}}^2$$
$$-{4\over 3}\left(F^a_{0i}\{\p_jA^a_i-gf^{abc}A_i^bA_j^c\}+(1/2)E_i^{a2}+E_i^a\p_i A_0^a+(u^a_i+\p_i c^{*a})\p_i c^a\right)\Big]\eqno(4.1)$$
where $F^a{\mu\nu}$ is defined in (A2), 
$C_G$ is the Casimir of the colour group, $c$ is the ghost field, and the source $u^a_i$ is introduced in (A1). The Feynman graphs and pseudo-Feynman graphs together have to have the right UV divergences
to be cancelled by these counter-terms. In Fig.4 we show some $O(g^4)$ graphs with counter-term insertions deduced from (4.1). For our present purposes, the only relevant term in (4.1) is
the second one.

\begin{figure}
\centering
\includegraphics[width=11cm]{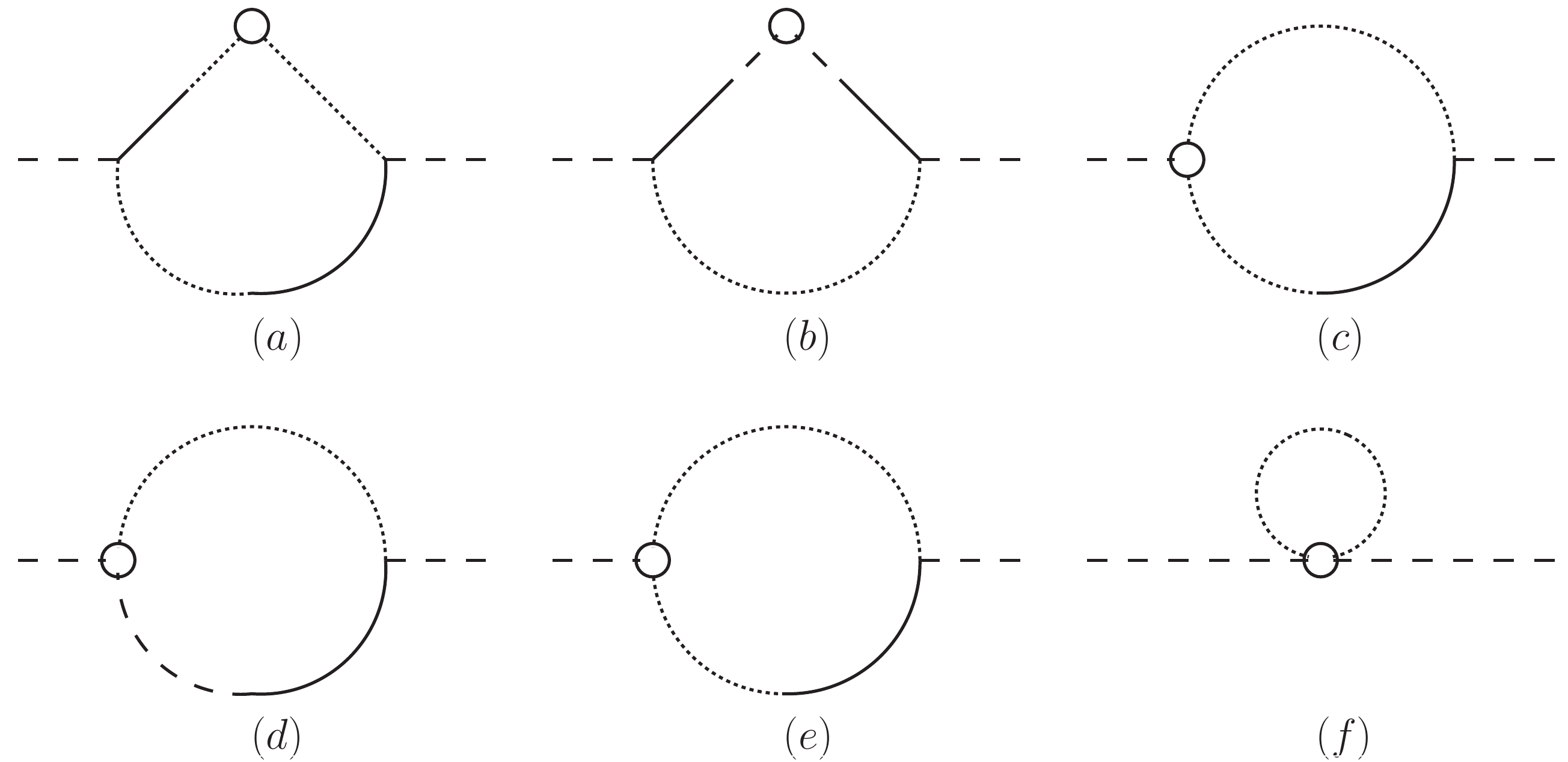}
\caption{ $O(g^4)$ graphs containing $O(g^2)$ counter-term insertions, which are represented by small circles.}
\end{figure}

We begin with Coulomb  self-energy sub-graphs. The 1-loop graphs which give the counter-terms
are shown in Fig.5(a) and (b). There is an important cancellation between these two graphs,
which ensures that their sum is proportional to $\K^2$, which is necessary in order to
cancel one of the two denominators from Coulomb propagators. Fig.6(c) is a 2-loop Feynman 
graph which contains Fig.5(b) as a sub-graph. Fig.6(b) is a pseudo-Feynman graph (equivalent to a contribution from the $V_2$ CL term) which contains Fig.5(a) as a sub-graph. Graph 6(b) comes with the
prescription that the $p_0$-integral is to be done first, then the $r_0$-integral; and this is
just what is required for renormalization, where divergent sub-integrations are to be done first.
\begin{figure}
\centering
\includegraphics[width=11cm]{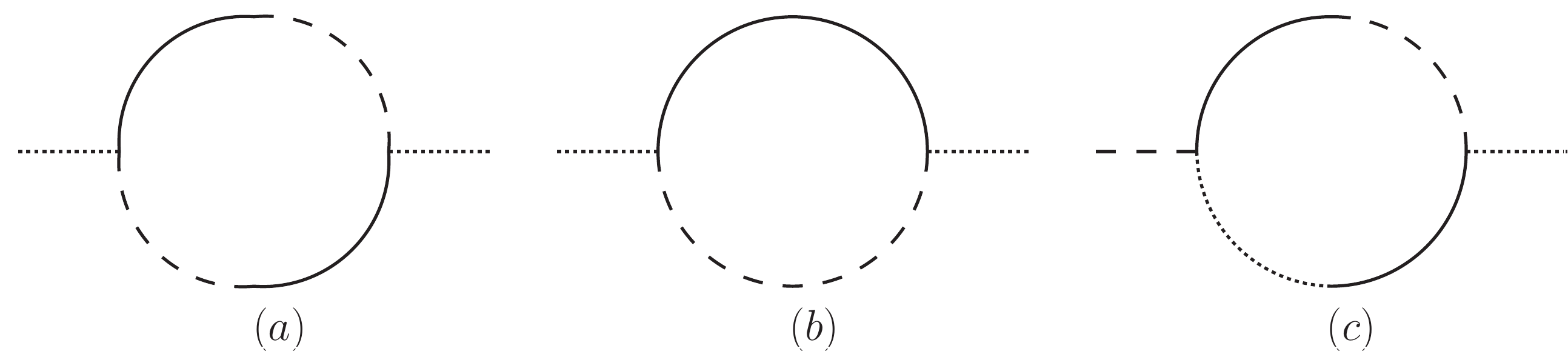}
\caption{  1-loop self-energy graphs. (a) and (b) are Coulomb line self-energy graphs.}
\end{figure}
\begin{figure}
\centering
\includegraphics[width=4in]{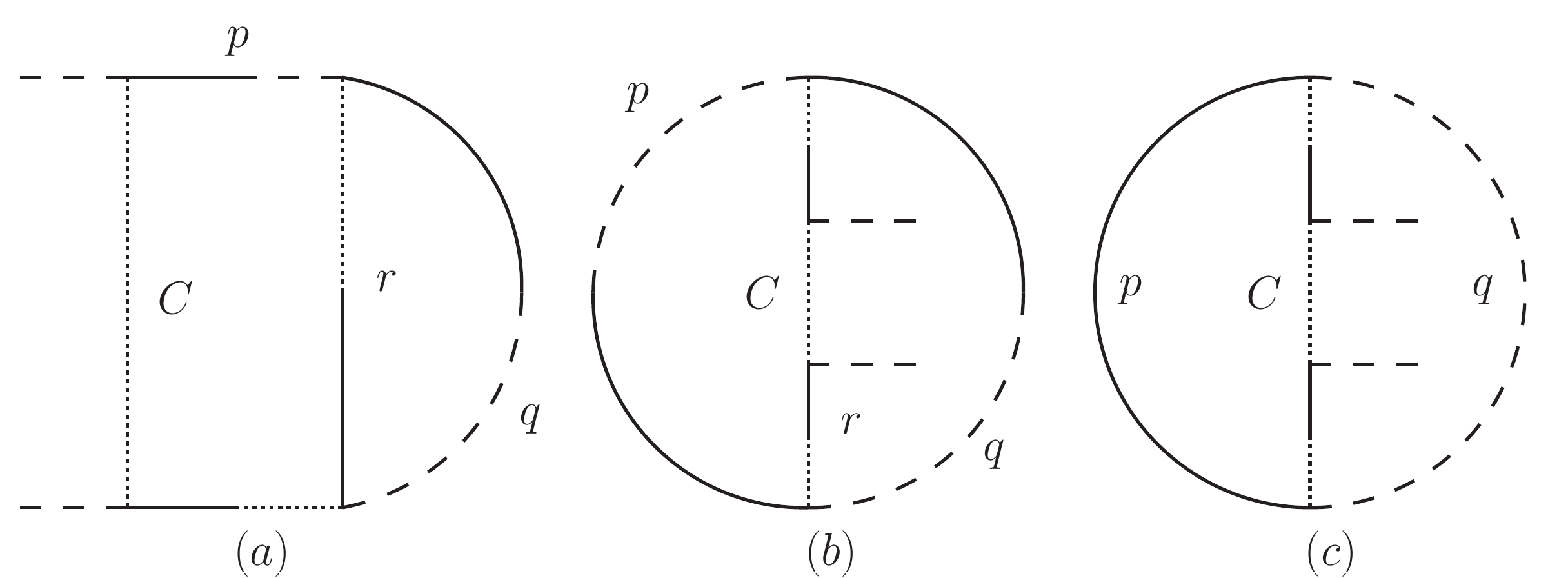}
\caption{$O(g^4)$ graphs containing self-energy subgraphs. (c) is an ordinary, unambiguous, Feynman graph. (b) is a pseudo- Feynman graph of the type of  Fig.2(a), and Fig.6(a) is of the type of Fig.3(b).
According to the prescription, in Fig.6(a) the integrals are to be done as in equation (1.5), giving zero; but in Fig.6(b) the integrals are to be done as in equation (1.3), that is
 integrating over the self-energy sub-graph  first.}
\end{figure}

Next we turn to UV divergent triangle sub-graphs. Fig.7 shows two such graphs. Fig.7(b)
is an ordinary, unambiguous Feynman graph, in which $p$, $p'$ and $q$ lines make up
a UV divergent triangle sub-graph. Fig.7(a) is a pseudo-Feynman graph,  of the class of Fig.2(a), also with a
UV divergent sub-graph triangle.
In Fig.7(a), the $p_0$-integral is to be done before the $r_0$-integral, and this is the order
needed for for the cancellation of the UV divergence (by the counter-term in Fig.4(e)).

\begin{figure}
\centering
\includegraphics[width=11cm]{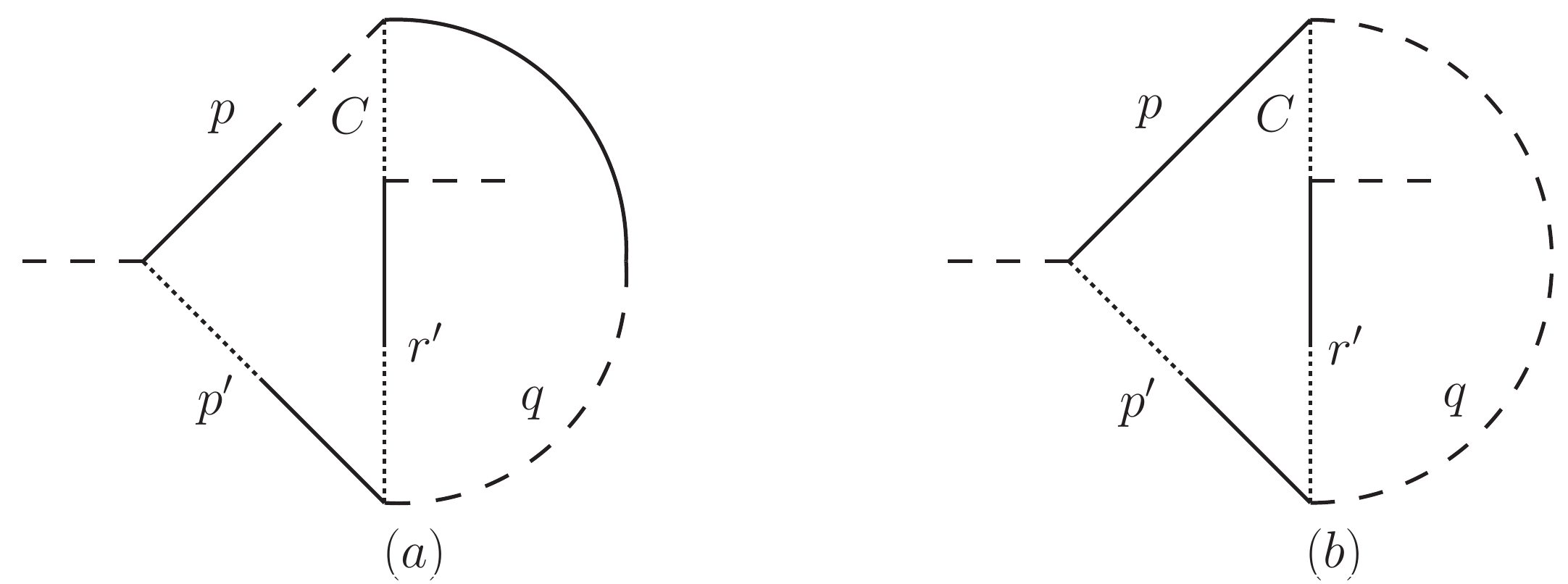}
\caption{ $O(g^4)$ graphs containing UV divergent triangle sub-graphs, made up of the lines $p,p',q$. (b) is an ordinary, unambiguous
Feynman graph. (a) is a pseudo-Feynman graph, of the class in Fig.2(a), derived from
a CL term. In (a), the $p_0$-integral is to be done before the $r_0$ one.}
\end{figure}

Fig.8 shows three more graphs, in which the triangle formed by the $p, q, q'$ lines is an UV divergent
sub-graph. Again Fig.8(b) is an ordinary unambiguous Feynman graph. Fig.8(a)
is of the class in Fig.2(b) and so gets zero contribution from $W_2$ in (2.11).
But there is a relevant graph Fig.8(c) of the class of Fig.3(b), coming from $W_1$ in (2.11).
The difference between (a) and (c) lies in a factor $P_iP_j/\P^2$ and because of the transversality
of the $q$ propagator this can be replaced by $P_iR'_j/\P^2$, and this contributes only an UV
finite part to the sub-integration. Thus the UV divergent sub-integrations in (b) and (c)
do combine as they should, and in fact cancel each other; so there is no need for a counter-term.

\begin{figure}
\centering
\includegraphics[width=11cm]{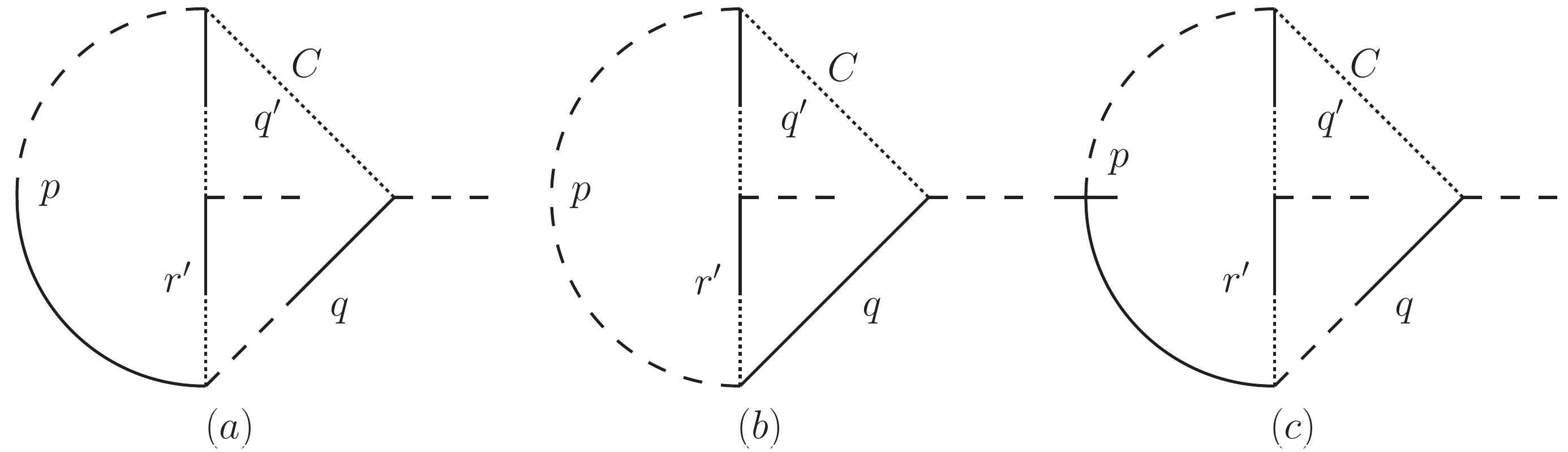}
\caption{$O(g^4)$ graphs containing UV divergent triangle sub-graphs made up of the lines $p,q,q'$. (b) is an ordinary, unambiguous
Feynman graph.  (a) is of the class in Fig.2(b) and so is zero. (c) is derived from $W_1$ in (2.11), and cancels the UV divergence in (b). In each graph, the symbol $C$ draws attention to the position of the
$G_0$ propagator in (2.3). The bar on the $p$-propagator in (c) indicated that only the $\delta_{ij}$
term in the numerator is retained.}
\end{figure}

We have verified  (at $O(g^4)$) the existence of sub-graphs with UV divergences necessary to 
cancel the counter-terms in Fig.4(a), (b), (c) and (e).  In the cases of (d) and (f), no ambiguous
graphs are involved.

\section{Energy divergences}

Another unfortunate feature of the Coulomb gauge is the presence, in individual graphs, of energy-divergences,
that is integrals of the form
$$\int dr_0 F(r_0, \R) \eqno(5.1)$$
where 
$$F\rightarrow F(\R)  \,\,\,\,\,\hbox{as}\,\,\,\,\,r_0\rightarrow \infty. \eqno(5.2)$$
Some, but not all, of these energy divergences occur in just the same graphs which have the ambiguous integrals
described in section 1, that is (to 2-loop order) graphs containing only two transverse gluon lines.
The ambiguous integrals show up when doing the integrals in the order
$$\int d^3\R d^3\P dr_0 dp_0.  \eqno(5.2)$$
The energy divergences appear from the order
$$\int d^3\R dr_0 d^3\P dp_0 \eqno(5.3)$$
which is the order required by renormalization theory.

It is expected that these energy divergences should cancel when sets of graphs are combined.  This has been
confirmed at $O(g^4)$ in \cite{AAdiv} but only  when the sub-graphs are
quark loops. Then, the $O(g^2)$ effective action is gauge-invariant, obeying Ward identities.
The analysis in \cite{AAdiv} made use of these Ward identities.
We want to extend the argument to cover gluon loop sub-graphs. Then the $O(g^2)$ effective
action is  in general not gauge-invariant, but only BRST invariant, so it is not obvious
that the cancellation of energy divergences is as simple. The divergent parts are shown in equation
(4.1), and indeed they are not all gauge invariant. However, the only one relevant
to the energy divergences is the second term in (4.1), which is gauge invariant.

The counter-term graphs in Fig.4 exhibit the energy divergences clearly, and the 2-loop
graphs whose sub-divergences Fig.4 cancel are the ones with energy divergences.

In order to extend the argument of \cite{AAdiv} to gluon loop sub-graphs, we need to show that these
sub-graphs at high energy obey Ward identities, not just BRST identities. The 
 divergent parts obey Ward identities  because the ghost graphs shown in Fig.9
 are UV convergent. In order to extend the argument to high-energy limits, we need to
 verify that the graphs like those Fig.9 are also suppressed at high energy. The graphs in Fig.9 have open ghost lines
 terminating at the ghost field $c^a$ and the  sources $u^a_0, u^a_i$ or $v^a_i$, which appear in the original Lagrangian in
 equation (A1) in Appendix.
 
 The graphs in Fig.9 all appear by power counting to be linearly divergent in the UV, but they
  are each proportional to $P_i$ (because the gluon line $k$ is transverse), and so are at most
  logarithmically divergent.  But in (b) there are three spatial indices, $i,j,k$ and so there must be
  another external spatial momentum. And there is  further cancellation between (c) and (d), which produces an  extra power of external spatial momenta. Thus (b) and (c)+(d)
  are actually UV convergent (as indicated by the absence of trilinear  ghost counter-terms in (4.1)).
  Also, to balance dimensions, we expect at least one power of $p_0$ in the denominator,
  suppressing the integrals at high-energy (that is $p_0\gg |\P|, |\Q|, q_0$). In Appendix B, we
  show this suppression in more detail.
  
  The above suppression does not occur in all ghost graphs. Fig.9(a) is UV convergent,
  but proportional only to one power of external spatial momenta ($Q_i$), and is in fact
  independent of $p_0$ and $q_0$. But, unlike  (b), (c) and (d),  graphs like (a) with $u^a_i$ or $u^a_0$ sources  do not enter into the
  BRST identities multiplied by a factor of energy, and so are irrelevant to the high-energy behaviour
  of the vertex parts.

 \begin{figure}
\centering
\includegraphics[width=11cm]{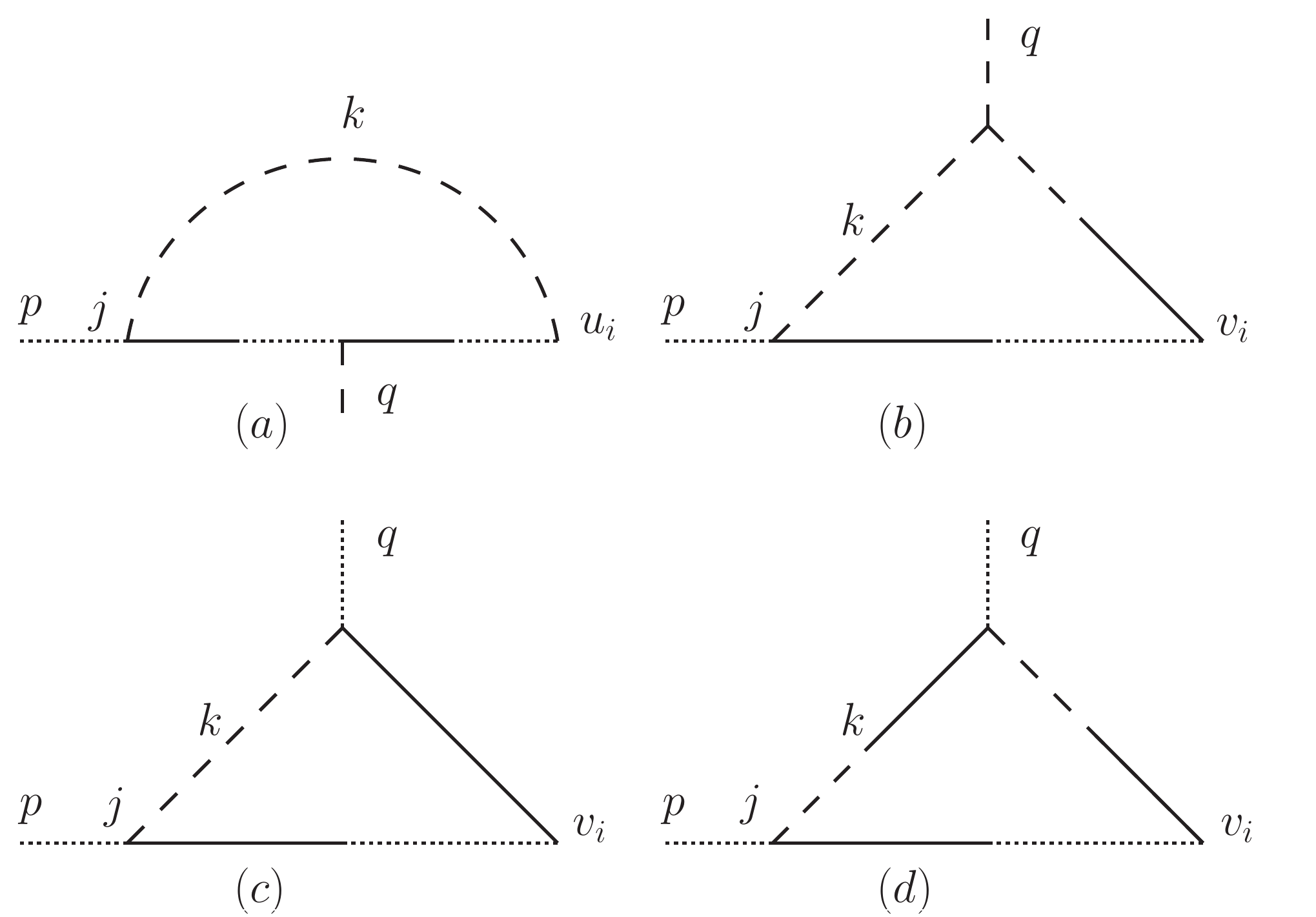}
\caption{Graphs contributing to the BRST identities. These graphs contain open ghost lines,
terminating at the Zinn-Justin source $u^a_0, u^a_i$ or $v^a_i$, and the bottom line is a ghost line. The integrals appear by power counting to be
linearly divergent in the UV, but they all develop external spatial momenta in their numerators
which cause them to be UV convergent. (b), (c) and (d) are energy-suppressed but the example in (a) is not.}
\end{figure}

\section{Conclusion}

To two loop order, the existence of the  Christ-Lee terms in the Hamiltonian for Coulomb gauge QCD
seems to be a problem for renormalization, because they do not have sub-graph structure of ordinary Feynman graphs. We have shown that this difficulty is overcome, by re-expressing the Christ-Lee
terms as pseudo-Feynamn graphs. We have also shown how energy divergences cancel
between Feynman and pseudo-Feynman graphs.

However, to three-loop order there may remain problems about the renormalization
of the Christ-Lee terms, when ultra-violet divergent sub-graphs are inserted into the ambiguous
2-loop graphs.  This question has been discussed in section 5(ii) of \cite{doust} and in \cite{doustjct}, \cite{AAjct}. 

This paper is only about perturbation theory. An important open question is whether there are
related complications in, for example. lattice calculations. Certainly, the Hamiltonian must be correctly
ordered, and this generates CL terms.

\section*{Appendix A - Feynman rules}

The Lagrangian density, in phase space form,  which we use is

$$ -{1\over 4}F^a_{ij}F^a_{ij}-{1\over 2}E^a_iE^a_i+E^a_iF^a_{0i}-{1\over 2\alpha}(\p_iA^a_i)^2$$
$$+(u^a_i+\p_i c^{a*})D_i^{ab}c^b
+u_0^aD_0^{ab}c^b-gf^{abc}v_i^aE_i^b c^c-{1\over 2}gf^{abc}K^a c^b c^c \eqno(A1)$$
where $\mu=(o,i)$, $c$ is the ghost field, $u^a_i, u^a_0, v^a_i$ and $K^a$ are sources. For $\alpha\neq 0$, the equations of motion are non-singular, and we can derive Feynman rules.  Then the limit
$\alpha \rightarrow 0$ is to be taken, to get the Feynman rules for the Coulomb gauge.
In (A1)
  $$F^a_{\mu\nu}=\p_{\mu}A^a_{\nu}-\p_{\nu}A^a_{\mu}+gf^{abc}A^b_{\mu}A^c_{\nu} \eqno(A2)$$
  and
  $$D^{ab}_{\mu}=\delta^{ab}\p_{\mu}-gf^{abc}A^c_{\mu} \eqno(A3)$$
  
  In (A1),  $A^a_0$ appears linearly and with no time derivative, so it may be integrated out;
 although we state Feynman rules in the form in which $A_0^a$ is retained. If it is integrated out, it
  enforces the Coulomb law
  $$D_i^{ab}E_i^b=0.\eqno(A4)$$
  From (2.1), this gives
  $$ E^a_i(\X)=E^{Ta}_i(\X)+\int d^3\Y G_i^{ab}(\X,\Y;\A)\rho^b(\Y)\eqno(A5)$$
  where $\p_i E^{Ta}_i=0$ and
  $$\rho^a=gf^{abc}A^b_iE^{Tc}_i. \eqno(A6)$$ 
   Using these equations and (2.4),
  $${1\over 2}\int d^3{\bf X}E_i^aE_i^a={1\over 2}\int d^2{\bf X}E_i^{Ta}E_i^{Ta}+{1\over 2}\int d^3{\bf  X}d^3{\bf X}'\rho^a({\bf X})C^{ab}({\bf X},{\bf X}';{\bf A})\rho^b({\bf X}') \eqno(A7)$$
  
  The second term in (A7) is the non-abelian Coulomb potential operator.   How to order the operators
  (which are all at the same time) in (A7) is one of  the questions which lead to the CL terms $V_1$ and $V_2$.

  The Feynman rules for (A1) are derived for instance in \cite{AAdiv}. They are shown in Fig.10 and Fig.11.
  In the Coulomb gauge, ghost lines are identical with parts of Coulomb lines. In fact closed ghost loops
  just cancel closed Coulomb loops. Therefore, only open ghost lines are required, where the
  ghost is attached to one of the sources $u_i^a, u^a_0, v^a_i, K^a$ which appear in (A1), and
  are shown in Fig.11.
  \begin{figure}
  \centering
  \includegraphics[width=7cm]{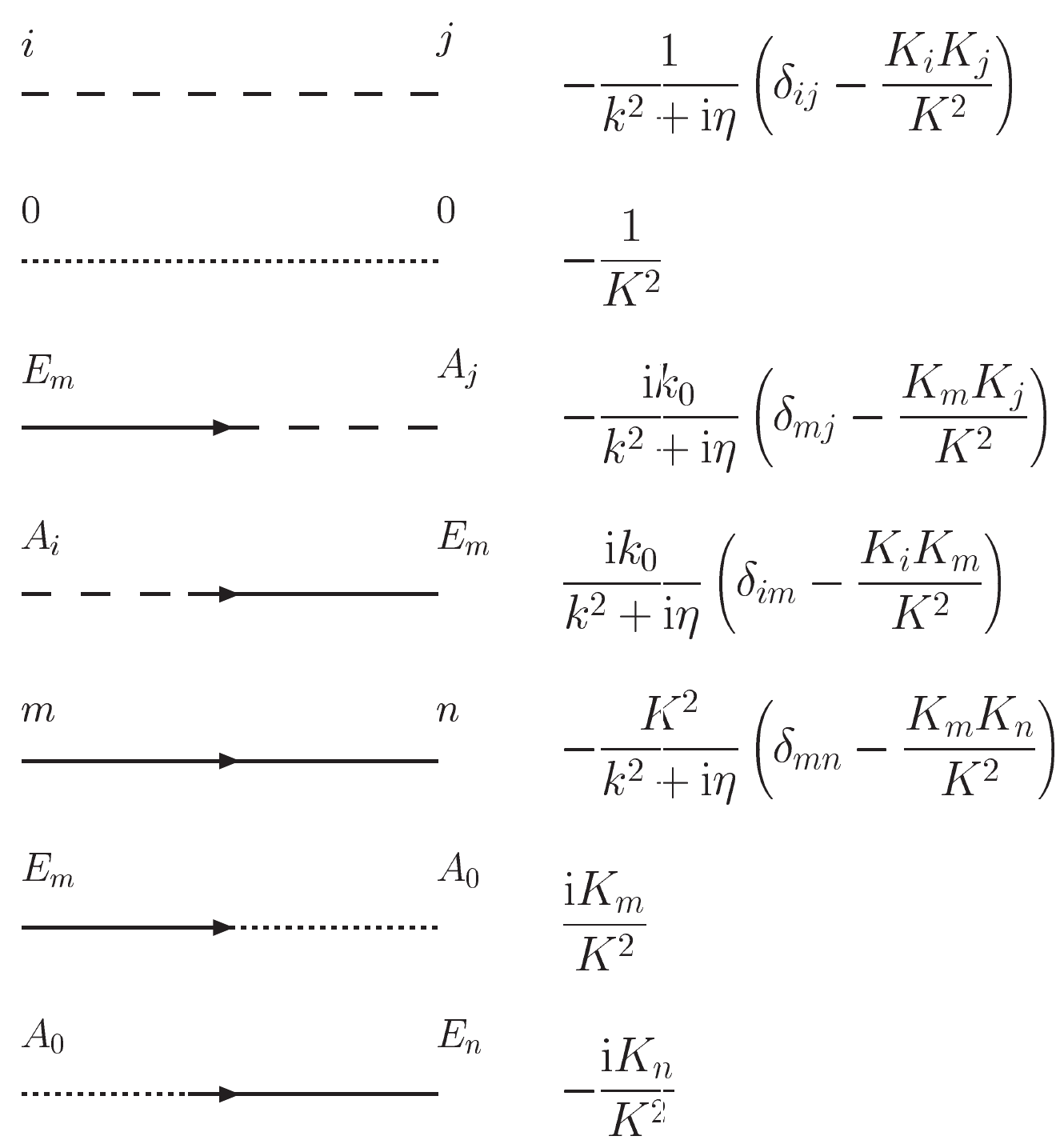}\\
  \includegraphics[width=7cm]{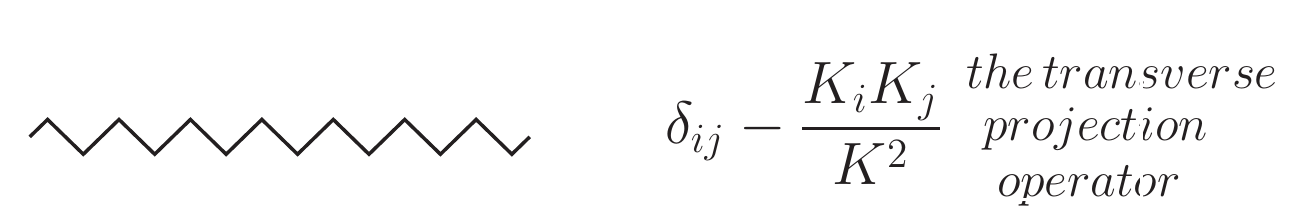}
  \caption{The Feynman rules for propagators. The general rule is that dashed lines represent
  $A^a_i$, dotted lines represent $A^a_0$ and continuous lines represent $E^a_i$.}
  \end{figure}
  \begin{figure}
\begin{center}
\includegraphics[width=9cm]{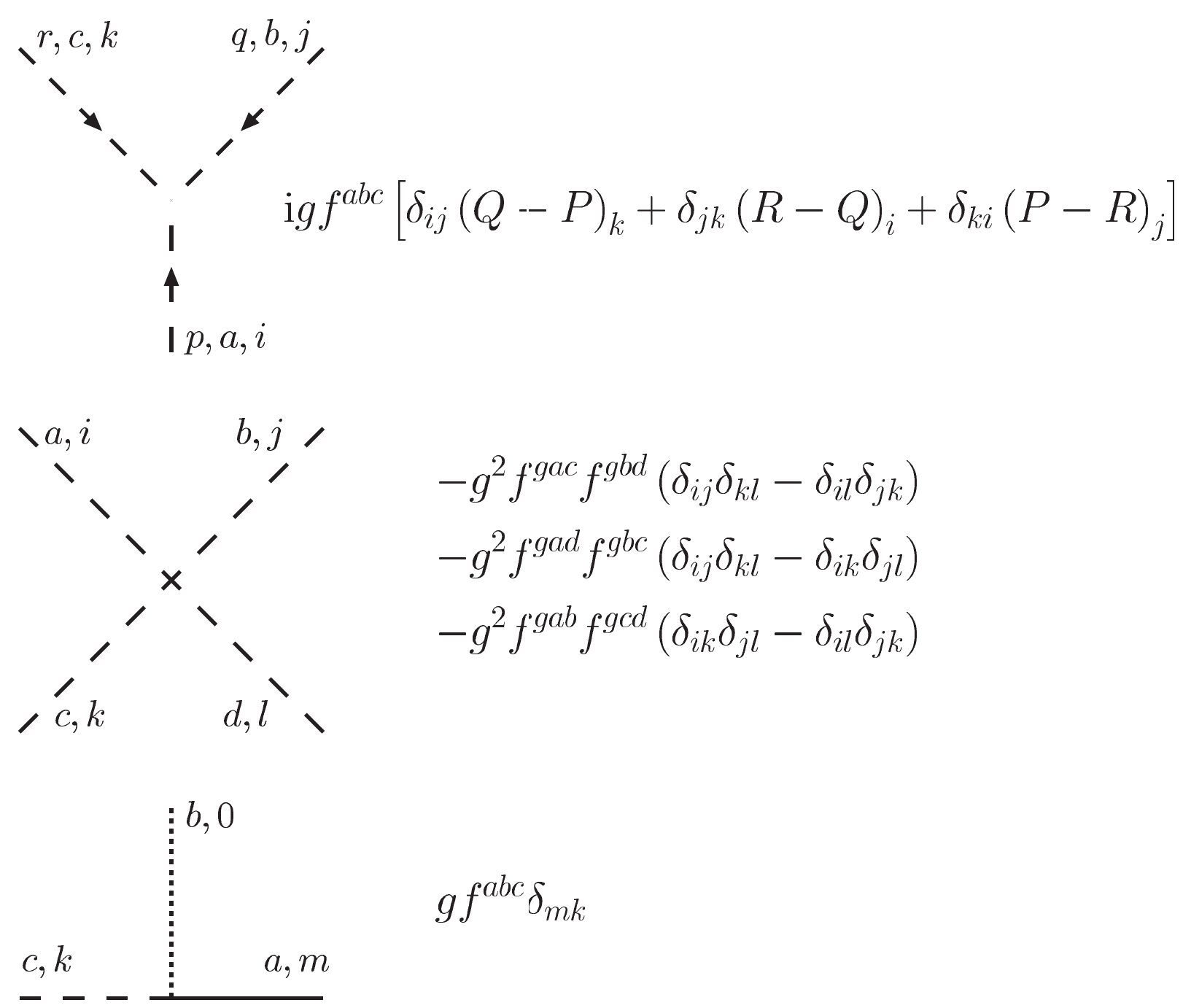}\\
~\\
~\\
\includegraphics[width=5.5cm]{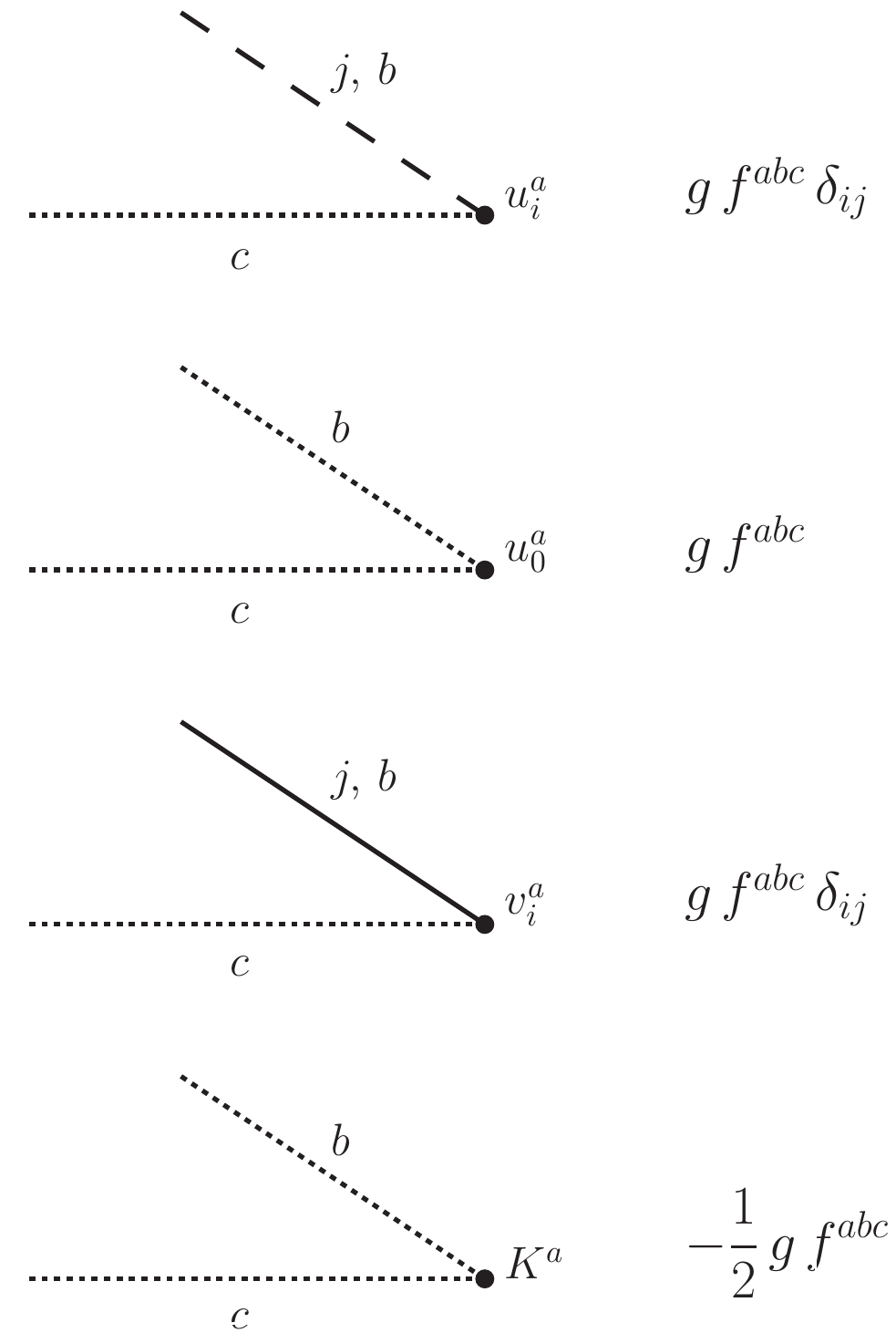}
\caption{Feynman rules for vertices.  The last four involve open ghost lines (with colour index  $c$) and ghost
sources.}

\end{center}
\end{figure}
\section*{Appendix B - High energy limits}

Here we examine the high energy limit of ghost Feynman graphs in more detail than in Section 5.
We take as an example Fig.9 (c) and (d). The sum of these two graphs contains the integral
$$P_j \int d^3\K dk_0{T_{jl}(\K)\over k_0^2-\K^2}{T_{li}(\K')\over {k'_0}^2-(\K')^2}{1\over (\K-\P)^2}\{{\K'}^2-k_0k'_0\}.\eqno(B1)$$
Doing the $k_0$ integral gives
$$i\pi P_j \int d^3\K {1 \over q_0-(|\K|+|\K'|)}{1\over (\K-\P)^2}\left( {|\K'|-|\K| \over |\K|} \right)T_{jl}(\K)T_{li}(\K')$$
$$+(q_0\rightarrow -q_0). \eqno(B2)$$

We are concerned with the limit
$q_0\gg |\P|,|\Q|$
and this comes from the region of integration $|\K|\gg |\P|,|\Q|$.  So the high-energy limit of (B2) is
$${i\pi P_j\over q_0}\int d^3\K{1\over (\K-\P)^2}\left( {|\K'|-|\K| \over |\K|} \right)T_{jl}(\K)T_{li}(\K')+(q_0\rightarrow -q_0)+o(1/q_0). \eqno(B3)$$
provided that this $\K$ integral is convergent. It is convergent because the factor $(|\K'|-|\K|)$
provides at least one power of $\Q$ and by rotational invariance there must be another factor
of $\P$ or $\Q$. So (B1) tends to zero faster than $1/q_0$ at high energy.

By a similar argument, the graph in Fig.9(b) is shown to have a high-energy limit of the form
$$ (1/q_0)Q_jM_{jli}(\P,\Q) \eqno(B4)$$
where $M$ is a third rank tensor with dimensions of a momentum..
\vskip 1cm
We are grateful to Dr. G. Duplan\v ci\' c for e-drawing the figures.


\section*{Bibliography}
\begin{enumerate}
\bibitem{zwanziger} D Zwanziger, Nucl. Phys. {\bf B518} 237 (1998)
\bibitem{cuch}A Cucchieri and D Zwanziger, Phys. Rev. {\bf D56}  014002 (2002)
\bibitem{scz} A P Szczepaniak and E S Swanton, Phys. Rev. {\bf D65}  025012 (2002)
\bibitem{szi} S Sziegel, G Krein and R S Marques de Carvalho, Brazilian J. Of Phys {\bf 34}  No.1 (2004)
\bibitem{alk} R Alkofer, M Kloker, A Krassnig and F Wagenburn, Phys. Tev. Lett. {\bf 96} 022001 (2006)
\bibitem{nak} A Nakamura and T Saito, Prog. Theor. Phys. {\bf 15}  189 (2005)
\bibitem{rei} H Reinhard and C Feuchter, Phys. Rev. {\bf D71}  105002 (2005)
\bibitem{lan} K Langfeld and L Moyaerts, Phys. Rev. {\bf D70}  074507 (2004)
\bibitem{epple} D Epple, H Reinhardt and W Schleifenbaum, Phys. Rev. {\bf D75} 045011 (2007)
\bibitem{burgio} G Burgio, M Quandt and H Reinhardt, Phys. Rev. Lett. {\bf 102} 032002 (2009)
\bibitem{camp} D R Campagnari, H Reinhardt and A Weber, Phys. Rev. {\bf D80} 025005 (2009)
\bibitem{rce} H Reinhardt,  D R Campagnari, D Epple, M Leder, M Pak and W Schleifenbaum, arXiv 0807.4635
\bibitem{naka} Y Nakagawa, A Voigt, E-M Ilgenfritz, M M\"{u}ller-Preussker, A Nakamura, T Saito, A Sternbeck and H Toki, arXiv:0902.4321 [hep-lat] (2009)
\bibitem{mohapatra}  R N Mohapatra, Phys. Rev {\bf D4}  378 and 1007 (1971)
\bibitem{christlee} N H Christ and T D Lee, Phys. Rev. {\bf D22}  939 (1980)
\bibitem{AAcanc} A Andra\v si and J C Taylor Eur. Phys. J C {\bf 41}  377 (2005)
\bibitem{AAdiv} A Andra\v si and J C Taylor, Annals of Physics {\bf 324}  2179 (2009)
\bibitem{cheng1} H Cheng and E-C Tsai, Chinese Journal of Physics {\bf 25}  95 (1987)
\bibitem{cheng2} H Cheng and E-C Tsai, Phys. Lett {\bf B176} (1986) 130, Phys. Rev. Lett. {\bf 57}  511 (1986)
\bibitem{doust}  Paul Doust, Annals of Physics {\bf 177}  169 (1987)
\bibitem{doustjct} P J Doust and J C Taylor, Physics Letters {\bf 197} 232 (1987)
\bibitem{AAjct} A Andra\v si and J C Taylor,  Annals of Physics {\bf 326} 1053 (2011)

\end{enumerate}
\end{document}